\documentclass[fleqn,usenatbib]{mnras}

\usepackage{newtxtext,newtxmath}
\usepackage[T1]{fontenc}
\usepackage{ae,aecompl}
\usepackage{placeins}
\usepackage{color,soul}
\usepackage{array}

\newcolumntype{P}[1]{>{\centering\arraybackslash}p{#1}}
\usepackage{graphicx}	% Including figure files
\usepackage{amsmath}	% Advanced maths commands
\usepackage{amssymb}	% Extra maths symbols
\usepackage{esvect}     % For vectors
\usepackage{enumitem} % Enumeration options

\title[Direct Wide-Field Radio Imaging]{Direct Wide-Field Radio Imaging in Real-Time at High Time Resolution using Antenna Electric Fields}

\author[J. Kent et al.]{
James Kent,$^{1}$\thanks{E-mail: jameschristopherkent@gmail.com}
Adam P. Beardsley,$^{2}$
Landman Bester,$^{3}$
Steve F. Gull,$^{1}$
\newauthor
Bojan Nikolic,$^{1}$
Jayce Dowell,$^{4}$
Nithyanandan Thyagarajan,$^{5}$\thanks{Nithyanandan Thyagarajan is a Jansky Fellow of the National Radio Astronomy Observatory.}
Greg B. Taylor,$^{4}$
\newauthor
Judd Bowman,$^{2}$
\\
$^{1}$Cavendish Laboratory, University of Cambridge, UK\\
$^{2}$School of Earth and Space Exploration, Arizona State University, Tempe, AZ, USA\\
$^{3}$Rhodes University, Drosty Rd, Grahamstown, 6139, South Africa \\
$^{4}$Department of Physics and Astronomy, University of New Mexico, Albuquerque, NM, USA\\
$^{5}$National Radio Astronomy Observatory, Socorro, NM, USA
}

\date{Accepted XXX. Received YYY; in original form ZZZ}

\pubyear{2019}
\begin{document}
\label{firstpage}
\pagerange{\pageref{firstpage}--\pageref{lastpage}}
\maketitle

\begin{abstract}

The recent demonstration of a real-time direct imaging radio interferometry correlator represents a new capability in 
radio astronomy. However wide field imaging with this method is challenging since wide-field
effects and array non-coplanarity degrade image quality if not compensated for. Here we present an alternative direct imaging correlation strategy using a Direct Fourier Transform (DFT), modelled
as a linear operator facilitating a matrix multiplication between the 
DFT matrix and a vector of the electric fields from each antenna. This offers perfect correction for wide field and non-coplanarity effects. When implemented with data from the Long Wavelength Array (LWA), it offers comparable computational performance to previously demonstrated direct imaging techniques, despite having a theoretically higher floating point cost. It also has additional
benefits, such as imaging sparse arrays and control over which sky co-ordinates are imaged, allowing variable pixel placement across an image. It is in practice a highly
flexible and efficient method of direct radio imaging when implemented on suitable arrays. A functioning Electric Field Direct imaging architecture using
the DFT is presented, alongside an exploration of techniques for
wide-field imaging similar to those in visibility based imaging,
and an explanation of why they do not fit well to imaging 
directly with the digitized electric field data. The DFT imaging
method is demonstrated on real data from the LWA 
telescope, alongside a detailed performance analysis, as well as an exploration of its 
applicability to other arrays. 
\end{abstract}


%%%%%%%%%%%%%%%%%%%%%%%%%%%%%%%%%%%%%%%%%%%%%%%%%%

%%%%%%%%%%%%%%%%% BODY OF PAPER %%%%%%%%%%%%%%%%%%

\section{Introduction}

The recent deployment of a functioning real-time direct imaging radio correlator, based on the E-Field Parallel Imaging Correlator (EPIC) \citep{thyagarajan_generic_2017} has demonstrated a new capability in radio interferometric imaging \citep{kent_real-time_2019}. 
This demonstration was performed on the Long Wavelength Array (LWA) \citep{taylor_first_2012-1},
using the Modular Optimal Frequency-Fourier (MOFF) mathematical formalism for direct Fourier imaging \citep{morales_enabling_2011} in the form of the EPIC correlator. 

The major benefit of using the MOFF formalism over a traditional FX correlator, where electric field measurements from different antennas are cross-correlated \citep{thompson_digital_2017},
is that it is able to reduce the computational scaling from $\mathcal{O}(n_a^2)$, to 
$\mathcal{O}(n_g\log{n_g})$, where $n_a$ is the number of antennas, and $n_g$ is the number of grid points in the aperture. This is because
direct imaging  does not involve having to compute a costly outer product operation, i.e., visibilities, with $\mathcal{O}(n_a^2)$ scaling,  between
the vector of channelised electric fields and its transpose, and provides
significant scaling benefits for dense arrays with large numbers of 
antenna elements. 

EPIC provides a capability for wide field of view imaging in real
time at high time resolution, unlocking a new capability in 
time domain radio astronomy. This can be used
for the investigation of Fast Radio Bursts (FRBs) on interferometers \citep{caleb_first_2017}, 
which have been recorded at low frequencies using instruments such as the Canadian
Hydrogen Intensity Mapping Experiment \citep[CHIME;][]{amiri_observations_2019}. 

In the original demonstration of the functioning EPIC correlator on the Sevilleta station of the LWA, 
non-coplanarity was not corrected for. Images were formed by gridding
the electric fields in a convolution step, and then applying a Fourier Transform using the
Fast Fourier Transform (FFT) algorithm \citep{thyagarajan_generic_2017}. The Sevilleta LWA site is a quite co-planar array, and at the low frequencies that the station operates at this does not lead to 
great distortion of the image as seen in the commissioning images in \cite{kent_real-time_2019},
demonstrating the high time resolution capacity. Various different approaches
have been used to solve for the $w$-term in the van-Cittert Zernike equation for
visibility based imaging, such as $w$-projection \citep{cornwell_noncoplanar_2008},
and $w$-stacking \citep{offringa_wsclean:_2014}.

For accurate wide field imaging, it is necessary to
correct for wide field effects arising from a non-coplanar measurement plane \citep{cornwell_radio-interferometric_1992}. As
a natural consequence of direct imaging, these non-coplanarities must be solved
and corrected for in real-time. Approaches to how to accomplish this vary, 
such as $w$-projection, where each visibility is convolved with a $w$-kernel
and then this convolved kernel is added to a regularly sampled grid \citep{cornwell_noncoplanar_2008}. Another method is $w$-stacking, where
the problem is split into layers along the $w$ co-ordinate, and a correction multiplied in at each layer \citep{offringa_wsclean:_2014}.

Using a $w$-projection and $w$-stacking method for solving for the non-coplanarities will be demonstrated and a cost analysis performed. However whilst the mathematics will be shown to be identical, correcting for wide-field effects using electric fields is practically difficult to do in real-time, which is necessary for direct imaging.

Finally, direct radio imaging using a Direct Fourier Transform (DFT) matrix with the digitized electric field data is shown to be a computationally tractable solution to the problem of real-time high time resolution wide field imaging with some additional attractive properties.
The DFT approach places no restrictions on the location of the antennas, thus allowing direct imaging of sparse arrays. 
It additionally gives fine grained control over the pixel locations on the sky.

This is reminiscent of beamforming techniques, where the antennas in the array are coherently summed into a beam to maximize the gain in a particular direction on the sky. This technique allows multiple phase centers for later correlation. It can also be used for reduction of data volume compared to correlating all antenna elements, by additively beamforming between a set of antennas and then correlating voltage beams from each beamformed set of antennas. This is known as Phased Array beamforming and can be used to reduce data rates and carefully sculpt beams in an interferometer, which is planned for the SKA-Low interferometer \citep{adami_ska_2011}. Beamforming can be done in both the voltage and frequency space, depending on the technique {\citep{barott_realtime_2011}}. Beamforming can be done as a dedicated observation mode for an antenna, for example in pulsar observations as at The Low Frequency Array (LOFAR) {\citep{jeannot_lofar_2011}}. It can also be used in concert with correlative techniques as described above. A related technique is the idea of tied array beamforming, as implemented on the Murchison Widefield Array (MWA) \citep{ord_mwa_2019} and LOFAR \citep{jeannot_lofar_2011}, where voltages are loaded from disc and beamformed to a few pixels on the sky, creating a set of steered voltage beams.

Beamforming can also be done using an FFT algorithm (with $\mathcal{O}(n_g\log{n_g})$ scaling) with a redundant layout of antennas to facilitate fast beamforming {\citep{masui_algorithms_2019}}.
This allows multiple antenna beams to be formed with different pointing angles, which can be monitored in real-time for transient detection, such as at CHIME \citep{amiri_observations_2019}. Examples of arrays that use a combination of these beamforming approaches include the LWA \citep{taylor_first_2012-1}, MWA \citep{tingay_murchison_2013,ord_mwa_2019}, CHIME \citep{stepp_canadian_2014}, and others.

The technique described here however is a direct implementation of the interferometry equation, which constitutes a correlation operation, which is multiplicative in nature, compared to beamforming which is additive. 
RThe DFT formalism facilitates a direct imaging correlator which can operate in real-time at high time resolution in exactly the same method as EPIC, but with the substitution of a grid and FFT step for a multiplication with a DFT matrix. 

Using a DFT matrix means that a sparsely distributed array can be used for direct electric field imaging, as it is released from the grid size constraint of the FFT algorithm to maintain real-time performance, such as in the EPIC correlator. Additionally any sky co-ordinates can be sampled at any resolution, thus allowing high pixel resolution images of the sky with a selectable field of view to be generated. True angular resolution is still limited by the dirty beam of the interferometer.

An overview of the theory dictating the MOFF formalism and wide field correction with electric field based imaging is shown in 
Section \ref{sec:theory}. An analysis of using a $w$-stacking technique
is shown in \ref{sec:wstacking} and shows why it is ultimately difficult to implement
in a practical direct imaging telescope. Using a Direct Fourier Transform
in real-time with data from the LWA at Sevilleta (LWA-SV) is shown in Section \ref{sec:dft}, along with a detailed performance analysis dissecting why the counter-intuitively high performance of the direct Fourier transform method is possible.

\section{Theory} \label{sec:theory}

Direct radio imaging, such as EPIC \citep{thyagarajan_generic_2017}, takes advantage of the multiplication 
convolution theorem to re-arrange the canonical van-Cittert Zernike theorem into a
Fourier relationship between the electric fields and the sky brightness distribution:
\begin{multline} \label{eq:vc2}
 I(l,m,w) = \Bigg\langle \bigg|\iint E(x,y,z) \times \\ \exp\big[2\pi i \big(xl + ym + z\big(\sqrt{1-l^2-m^2}-1\big)\big)\big]dx\: dy \bigg|^2 \Bigg \rangle \mbox{.}
\end{multline}

Where the electric fields are measured at antenna locations in $(x,y,z)$, which are the physical locations of the antennas in the local co-ordinate frame of reference in units of wavelengths at the sampled frequency. The electric fields are convolved with the antenna illumination pattern onto a grid at this location. Then a Fourier transform is performed,
the resulting matrix is squared by its complex conjugate, and accumulated over $N_{T}$ timestamps.  This is exactly the same as a ``dirty" image formed with visibilities. 

The key difference is that the electric fields are measured in the $x,y,z$ system of antenna physical locations, whereas with visibilities they are measured in $u,v,w$ which is a vector projected along the baseline between two antennas. Both co-ordinate systems have the same basis vectors. The result, $I(l,m,w)$ is the same, so $z$ in this electric field frame of reference is of the same set of basis vectors as $w$ in the visibility frame. Henceforth we will refer to this non-coplanarity dictated by $z$/$w$ in terms of $w$, for harmony with existing literature.

The above equation has the same  non-coplanarity $w$-term \citep{cornwell_radio-interferometric_1992} as exists with visibility based imaging. This 
intuitively makes sense as the visibilities are the cross-correlations of the 
electric fields, represented as an outer product of the vector of electric fields
and their complex conjugates:
\begin{align} \label{eq:vis}
  V_{12}(u,v,w) = \big\langle \mathbfit{E}_{1}(x_1,y_1,z_1) \otimes \mathbfit{E}_{2}(x_2,y_2,z_2)^{\ast} \mbox{.} \big\rangle
\end{align}

If we go further we can show that the electric field contributions from the sky can be
modelled as: 
\begin{align} \label{eq:efieldm}
    E(r) = \int E(\Omega_k) e ^ {-i \phi } e ^ {-i\mathbfit{k}\cdot \mathbfit{r}} d\Omega_k
\end{align}
and visibilities as:
\begin{align} \label{eq:efieldmv}
    V(r) = \int I(\Omega_k)e ^ {-i\mathbfit{k}\cdot \mathbfit{r}} d\Omega_k \mbox{,}
\end{align}
where $\mathbfit{k}$ is a vector of sky cosine co-ordinates and $\mathbfit{r}$ a vector of measurement plane co-ordinates. $E(\Omega_k)$ and $I(\Omega_k)$ represent the electric field and intensity pattern respectively at a particular location on the sky. We integrate over the infinitesimal solid angles $d\Omega_k$.
$e^{-i\phi}$ is a random phase term indicating that all points on the complex sky are mostly incoherent with respect to each other.

Both of these equations satisfy the Helmholtz equation, thus constitute a valid wave equation. This property along with Equation \ref{eq:vc2} suggests that the same approaches to wide field correction should apply for electric fields due to them allowing the same classes of solutions.  Thus any valid method for $w$-correction with visibilities, might  also work with electric fields due to them using the same set of basis functions.

With the above relations in mind, multiple techniques can be used to correct for non-coplanarity, such as $w$-projection \citep{cornwell_noncoplanar_2008}, 
$w$-stacking \citep{offringa_wsclean:_2014}, or optimal gridding functions \citep{ye_optimal_2019}.

\subsection{Direct Fourier Transform Operator}

Using the direct Fourier transform is by far the easiest method, but suffers from poor 
scaling as the image size increases. But this is still significantly better for electric
fields than visibilities. With electric fields, the scaling is $\mathcal{O}(N_K N_A)$,
compared to $\mathcal{O}(N_K N_A^2)$ with visibilities. $N_{K}$ is the number of sky
pixels ($l$,$m$ co-ordinates), and $N_{A}$ is the number of antennas.

The Fourier relationship from the electric fields can be viewed as a bilinear map from the electric fields to the dirty map space, where the dirty map is the true sky
convolved with the dirty beam of the instrument:
\begin{align} \label{eq:dftop}
    S = \big|DX\big|^2 \mbox{,}
\end{align}

where we define $S$ as the real matrix representing the sky-modes sampled at a discrete
set of sky cosine co-ordinates. $D$ is the complex DFT matrix representing the direct Fourier transform
of the matrix $X$ which is our electric field data matrix. The absolute value squared of $S$ in Equation \ref{eq:dftop} indicates taking the magnitude of each complex entry in the $DX$ matrix and squaring it. This is equivalent to a Hadamard product between $DX$ and its complex conjugate $(DX)^\ast$. The DFT matrix $D$ is of the form:
\begin{equation} \label{eq:dft_matrix}
	D = \frac{1}{N}\begin{bmatrix} 
	\omega_{0,0} &  \omega_{0,j} & \dots  & \omega_{0,N_A} \\
	\omega_{i,0} &  \ddots & & \omega_{i,N_A} \\
	\vdots &  & \ddots & \vdots \\
	\omega_{N_K,0} & \dots &  & \omega_{N_K,N_A} \\
	\end{bmatrix} \mbox{,}
\end{equation}
			
where $w_{i,j} = exp \left[ -i2 \pi k_i \cdot r_j \right] $, with $i$ representing
the index of the sky cosine co-ordinate being sampled, and $j$ the index of the antenna
from which the electric fields are being sampled. 

$S$ has dimensions of $N_{K}$ rows and $N_{T}$ columns, where $N_{T}$ is the number of timestamps being imaged. The $D$ matrix is time-independent and thus can be pre-computed for different observations. Its dimensions are $N_{K}$ rows and $N_{A}$ columns. The $X$ matrix has dimensions of $N_{A}$ rows and $N_{T}$ columns. This allows batch imaging of multiple timestamps. 

One of the bonuses of using the DFT over an FFT is that there is  flexibility in which sky pixels are sampled due to not being held to the requirement of a regular grid. It also places
no limitations on the placement of the antennas in the measurement plane, whereas with the FFT the limitation is the finite grid size, and increasing this grid size increases computational cost. 

Thus the DFT allows sparse
arrays to be imaged using the EPIC correlator. By re-generating the $D$ matrix during an observation, therefore explicitly adding
time dependence, different observation modes can be incorporated such as tracking celestial objects and imaging them at high time cadence and high resolution. 

Within the original EPIC architecture described by \cite{thyagarajan_generic_2017}, it is shown that EPIC is a generic framework that allows for optimal image making with 
heterogenous arrays. This is where the antenna's have different properties such as:
\begin{enumerate}[leftmargin=0.5cm]
    \item Cable Complex Gains
    \item Antenna Complex Gains
    \item Antenna Illumination Pattern
\end{enumerate}

These are still able to be dealt with using the DFT operator. 
The operator $D$ in Equation \ref{eq:dftop} can have these terms
folded into them. For example take (i). The complex gains from the cables should be 
known from the characterisation of the instrument, and are 
direction independent. Thus the correct gains and phases can be applied to the $D$ matrix. This can
be modelled as a Hadamard product between a matrix $A_{c}$ and
$D$:
\begin{align}
    \bar{D} = D \odot A_{c} \mbox{.}
\end{align}

Where the rows of $A_{c}$ are identical in each column,
but the columns differ, corresponding to the individual antenna's
gains and phases. Next moving to (ii) this is also a position 
independent term but naturally has a time dependence associated with 
it and must be solved through calibration of the system. An 
example of calibrating an electric field based direct imager
has been demonstrated by \cite{beardsley_efficient_2017}, 
and this can be folded into a separate  complex $A_{s}$ matrix, 
with the caveat
that there is now a time dependence as the calibration solutions
naturally change over time, and this is also multiplied point-wise
with the above:
\begin{align}
    \bar{D} = D \odot A_{c} \odot A_{s} \mbox{.}
\end{align}

Taking into account (iii) is slightly more difficult due its dependence on position and antenna. 
Thus there
will be another matrix defined, $A_{i}$ representing the 
electric field patterns, the Fourier transform of the 
antenna illumination pattern, at each sky cosine 
co-ordinate and antenna. Again we can do another 
point-wise multiplication:
\begin{align}
    \bar{D} = D \odot A_{c} \odot A_{s} \odot A_{i} \mbox{.}
\end{align}

Thus Equation \ref{eq:dftop} becomes:
\begin{align} \label{eq:dftop_full}
    S = \big|\big(D \odot A_{c} \odot A_{s} \odot A_{i}\big)X\big|^2 \mbox{.}
\end{align}

The antenna beam correction is done in the sky space compared to in the 
measurement plane as described in \cite{thyagarajan_generic_2017}, using the multipliacation-convolution theorem, which is mathematically equivalent to techniques such as A$w$-projection \citep{bhatnagar_correcting_2008}. 

Thus the DFT approach is equivalent to the MOFF formalism shown in \cite{morales_enabling_2011}, with
several key differences. The major one is that the scaling 
is no longer $\mathcal{O}(n_g\log{n_g})$ due to not 
convolving onto a regular grid and using the FFT, which is the architecture used previously \citep{thyagarajan_generic_2017}. It is now
$\mathcal{O}(N_K N_T N_A)$, thus for arrays with many
antennas or producing images with many different
sky positions sampled the cost increases linearly for each
dimension. This will be demonstrated to be computationally
fast enough to run a real-time direct radio imager using 
data from  LWA-SV.

The benefits of this approach however are many, including that
one is no longer constrained to using a small dense array 
as is the case with the original EPIC formulation. The antennas can be located anywhere (with additional consideration required for 
ionospheric behaviour), and the sky can be sampled at any
location. Thus it is feasible to make high time resolution
images of the sky using the electric fields directly by 
taking advantage of the speed of the matrix multiplication
in Equation \ref{eq:dftop} on modern GPU hardware.

\section{w-Projection and w-Stacking} \label{sec:wstacking}

Before consideration of directly solving non-coplanarity through
the use of the DFT, existing schemes for solving non-coplanarity (the $w$-term) with visibilities were explored to understand if
they are applicable to direct imaging using the E-Fields. To this end, $w$-projection \citep{cornwell_noncoplanar_2008} and $w$-stacking \citep{offringa_wsclean:_2014}, were
chosen. These methods were explored and tested in detail to explore whether they are both mathematically capable of correcting for the $w$-term with
electric fields, as well as if they are practically efficient to 
implement. 

In the original formulation of EPIC, a convolution was performed 
which mapped the electric fields to the measurement plane, and
then an inverse Fourier transform was applied using the FFT algorithm.
To apply $w$-correction in this method, we could convolve the electric
fields with a Fourier transformed Fresnel pattern, the $w$-kernel, in the same way 
that $w$-projection works \citep{cornwell_noncoplanar_2008}. Unfortunately
the grid sizes in EPIC are often small to account for dense arrays and to constrain the computational cost of the FFT. This means that the size of convolutional kernel is limited by the grid size. The size of the 
$w$-kernel can be determined as in \cite{mitchell_science_2014}:

\begin{align}
    \mbox{Kernel Size(1-D)} = \Bigg[ \bigg[ \frac{w_{max} \theta}{2} \bigg]^2 + \bigg[ \frac{w_{max}^{\frac{3}{2}} \theta}{2 \pi \epsilon} \bigg] \Bigg]^{\frac{1}{2}} \mbox{,}
\end{align}

$\theta$ represents the field of view size, $w_{max}$ the maximum $w$ value, and $\epsilon$ the fraction of peak to represent the $w$ pattern out to. With a field of view set to the entire sky, $\epsilon$ set to 0.01, and $w_{max}$ set to 10, similar to non-coplanarity in LWA-SV, this results in a recommended convolution size of more than 40, which is impractical for small grid sizes used at low frequencies, such as those used in the implementation of the EPIC correlator at LWA-SV \citep{kent_real-time_2019}. This would result in having to increase the grid size to account for the convolution, with a commensurate increase in computational
cost, resulting in compromises having to be made on the number of channels that can be processed simultaenously. However at higher frequencies where grid sizes increase in the EPIC correlator (see \citealt{thyagarajan_generic_2017}) it may be a more tractable solution to the problem of wide field imaging. 

As a convolution based approach to correcting the $w$-term would
necessarily increase computational cost due larger grid sizes, it may be beneficial to 
look at approaches to correcting the $w$-term in sky space. This is the approach followed 
in WSClean \citep{offringa_wsclean:_2014} using $w$-stacking, where we grid each electric field at a particular ``plane",
where the planes are spaced out in $w$, creating a stack. Then a $w$-correction, a Fresnel pattern, is multiplied 
in between ``plane", and the planes are iterated through until all electric field values are gridded.

The explicit steps for correction in this way are:

\begin{enumerate}[leftmargin=0.5cm,label=(\arabic*),style=sameline]
    \item Sort electric fields in order of increasing/decreasing $w$.
    \item Apply electric field measurement to current $w$-plane, if it is the nearest one. Optionally use anti-aliasing kernel to increase accuracy. 
    \item Execute inverse FFT to image plane
    \item Multiply Fresnel pattern to plane.
    \item Execute forward FFT to measurement plane.
    \item Repeat steps (2) through (5) for all planes.
    \item Transfer plane back to w=0 plane. Output image.  
\end{enumerate}

The degree of correction for the $w$-term in Equation \ref{eq:vis} is contingent
on the spacing of the $w$-planes. A $w$-projection kernel can also be applied to decrease
the number of $w$-planes. 

To validate that this mathematically works well, electric fields
were simulated using Equation \ref{eq:efieldm} for a series of
point sources. Each source's phase was randomised compared to the others to ensure
they are incoherent. They were then sampled at a set of discrete locations
in the measurement volume, corresponding to a 3-D Gaussian distribution of points. Whilst
no interferometer would look like this in practice, it allows us to show that it is still
possible to make correct images by applying $w$-stacking to the electric fields, even 
in this artificial worst case.

\begin{figure}
    \centering
    \includegraphics[width=1.0\columnwidth]{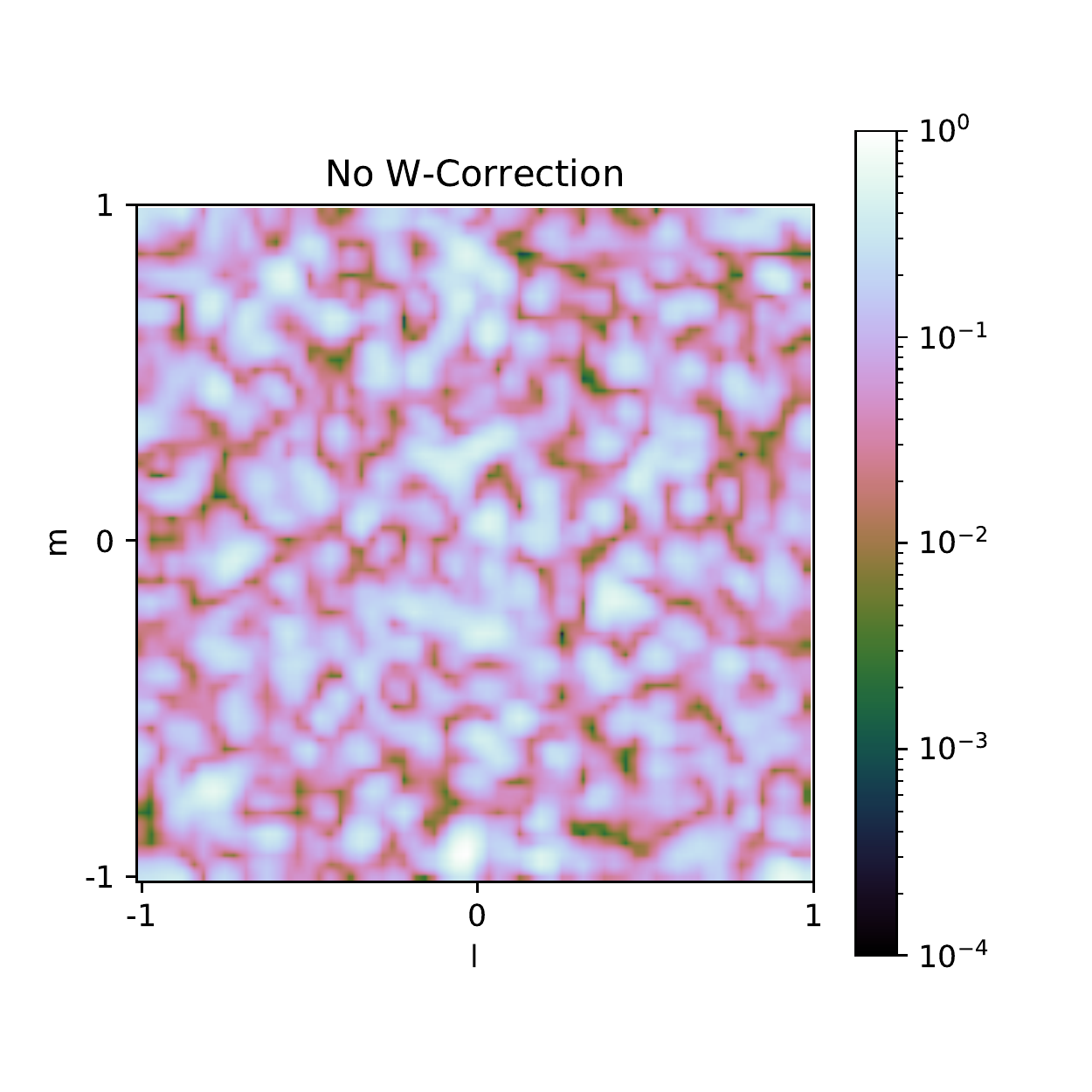}
    \includegraphics[width=1.0\columnwidth]{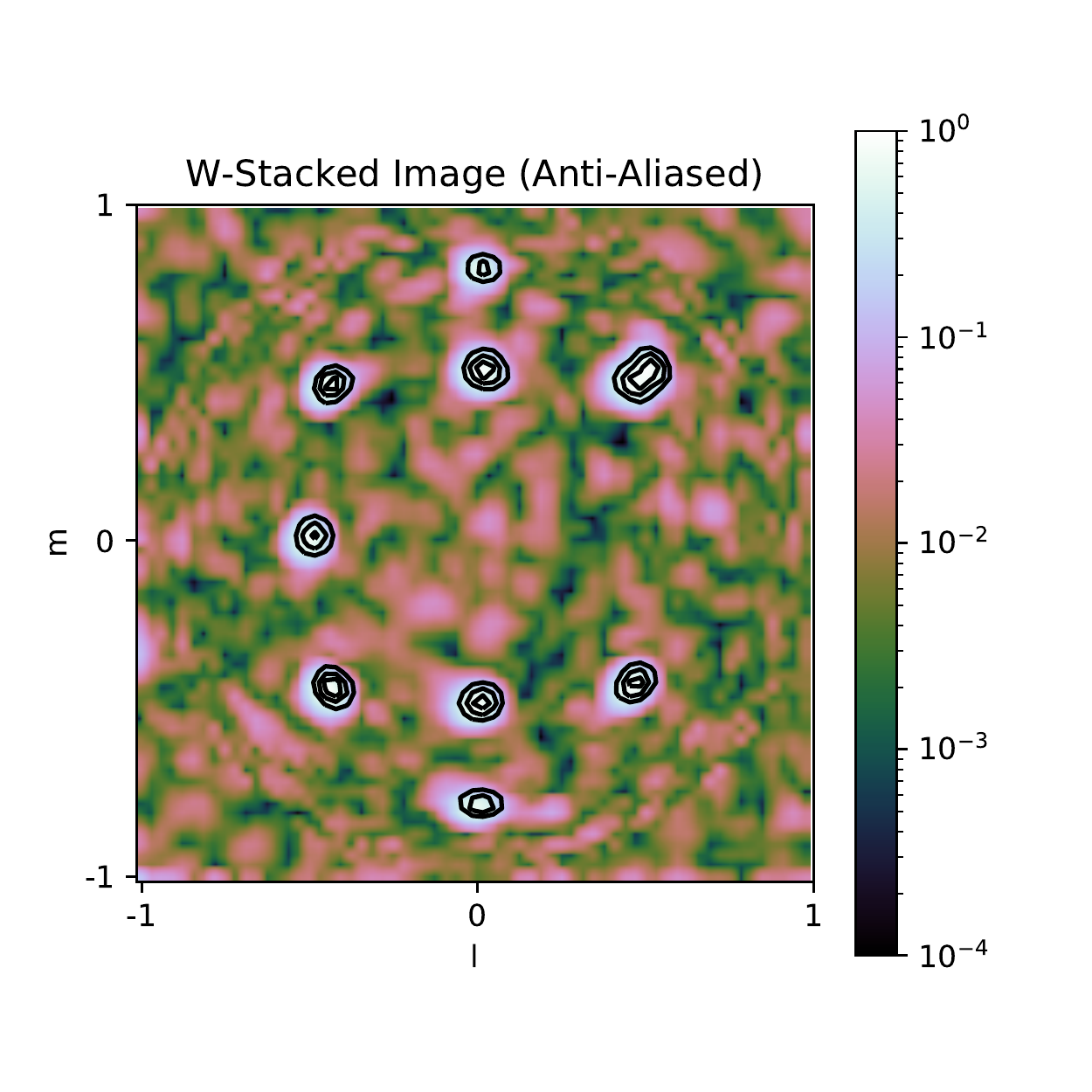}
    %0.5\textwidth]{images/time_gulp_scaling.pdf}
    \caption{(top) The sky imaged without any $w$-correction. There is no realistic sky brightness distribution recovered. Pixel values are in arbitrary units. (bottom)A simulated  sky imaged with $w$-correction using direct $w$-stacking. Our point sources are clearly imaged. Pixel values are in arbitrary units. The sky has been normalised to 1. Contours start at 0 and correspond to levels of 0.2.}
    \label{fig:correction}
\end{figure}

The top image in Figure \ref{fig:correction} is the sky brightness distribution
without any $w$-correction applied. The image is completely incoherent and 
wholly unrepresentative of the true sky intensity distribution for the dirty map.

\begin{figure}
    \centering
    \includegraphics[width=1.0\columnwidth]{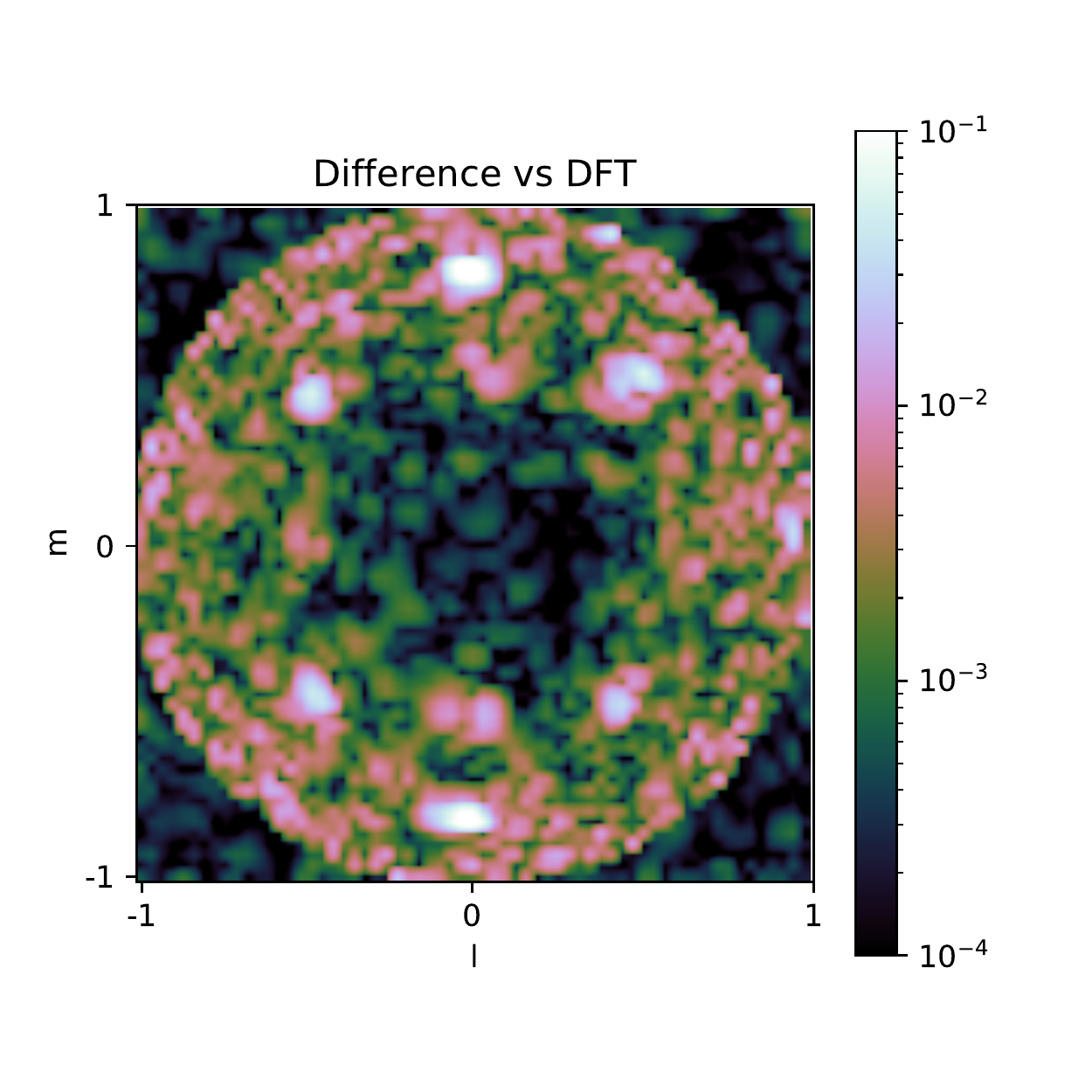}
    %0.5\textwidth]{images/time_gulp_scaling.pdf}
    \caption{A difference image between the image in Figure \ref{fig:correction} and an image created using the direct Fourier transform, with perfect $w$-correction. A Prolate Spheroidal Wave Function(PSWF) anti-aliasing function is applied to the electric fields to account for sub-grid point sampling. Thus the equivalency of $w$-correction between E-Fields and visibilities is shown. However $w$-stacking with electric fields is not computationally feasible in real-time.}
    \label{fig:diffvsdft}
\end{figure}

After $w$-correction using $w$-stacking, the sky brightness distribution shown 
in the bottom plot in Figure \ref{fig:correction} is recovered. The stacks in this case are calculated every $w$=0.1. An anti-aliasing kernel is also applied to the image, using a prolate spheroidal wave function 
\citep{jackson_selection_1991}.
A difference image is formed versus an image calculated using the direct Fourier
transform implementation of Equation \ref{eq:vc2}, and this is shown in 
Figure \ref{fig:diffvsdft}. The dirty image is recovered to an error of 1 part in $10^4$ averaged across 
the image, with pixels corresponding to points having slightly higher errors of between 1 part in $10^2$ and 
1 part in $10^3$. 

This demonstrates that wide field correction can be performed in the same way as visibilities using the electric fields, however this is not an efficient method in practice. $w$-projection may be a good method with large grid sizes, but this is likely to not be practical until consumer computing hardware increases in power to allow an EPIC correlator implemented on a higher frequency interferometer with commensurately larger grid sizes, and the $w$-projection overhead. With $w$-stacking, there is the requirement that these stacks be processed at every single time iteration of the electric field measurements. This is wholly impractical
with current computing hardware, however it is a useful result to know what the similar mathematical schemes apply as with visibilities. The need to do wide-field correction in real-time with direct imaging provides a very difficult constraint to performing this technique on a working interferometer.

\begin{figure}
    \centering
    \includegraphics[width=1.0\columnwidth]{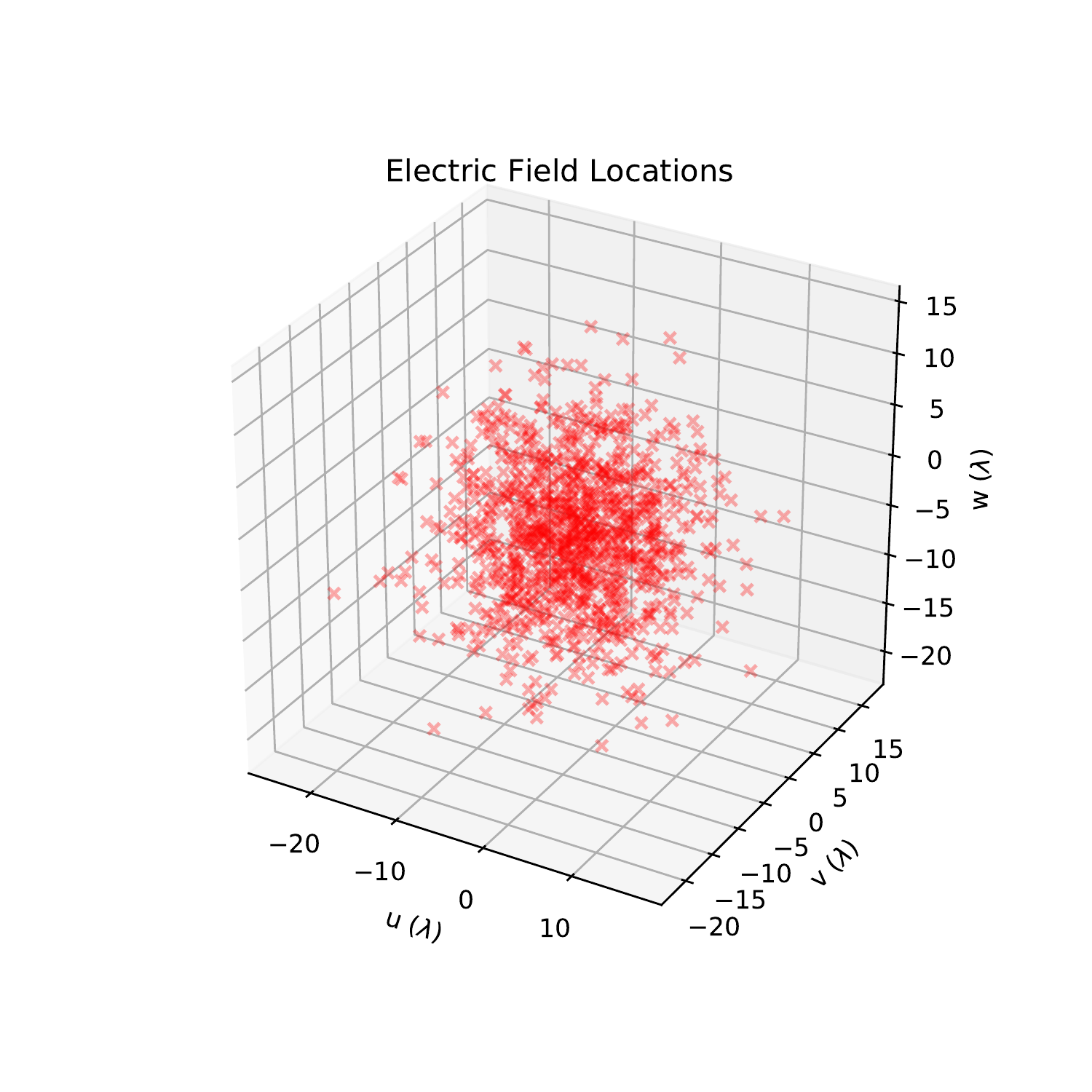}
    %0.5\textwidth]{images/time_gulp_scaling.pdf}
    \caption{Simulated antenna locations for electric field measurement. Making a worst case, practically impossible distribution of points for co-planarity allows us to demonstrate that it can still be mathematically corrected. }
    \label{fig:elocs}
\end{figure}

\section{Direct Fourier Transform Imaging} \label{sec:dft}

Using a direct Fourier transform is an attractive alternative 
mathematically because of its perfect $w$-correction. However in 
practice a DFT is often considered computationally unfeasible, but
using the DFT formalism described earlier it can be written as a 
dense matrix multiplication of the electric fields with a DFT
matrix. This approach
was tested on datasets from the LWA to explore its performance,
and a broader analysis was undertaken to understand its practicality
, and applicability to other arrays.

\subsection{The Long Wavelength Array}

The direct Fourier transform method was tested using 
pre-captured electric field data from the LWA. The original direct imaging pipeline described in \cite{kent_real-time_2019}, was 
modified to facilitate this. The original and modified
pipeline were both implemented using the Bifrost framework \citep{cranmer_bifrost:_2017-1}\footnote{The source code for the original EPIC correlator pipeline for the LWA, and the DFT pipeline is available at: \url{https://github.com/epic-astronomy/LWA_EPIC}.}.

The LWA is an interferometer currently located at two 
sites in New Mexico, USA. The site used for our analysis
here is the Sevilleta site, LWA-SV, which is the same one used
for the demonstration of the EPIC correlator in \cite{kent_real-time_2019}. The LWA operates between
frequencies of 10 and 88 MHz. Each site consists of 
256 dual orthogonal polarization dipole antennas with a wide beam.
The array is organised
into a dense central core of pseudo-randomly located antennas,
with an outrigger antenna providing greater angular 
resolution.

\begin{figure}
    \centering
    \includegraphics[width=1.2\columnwidth]{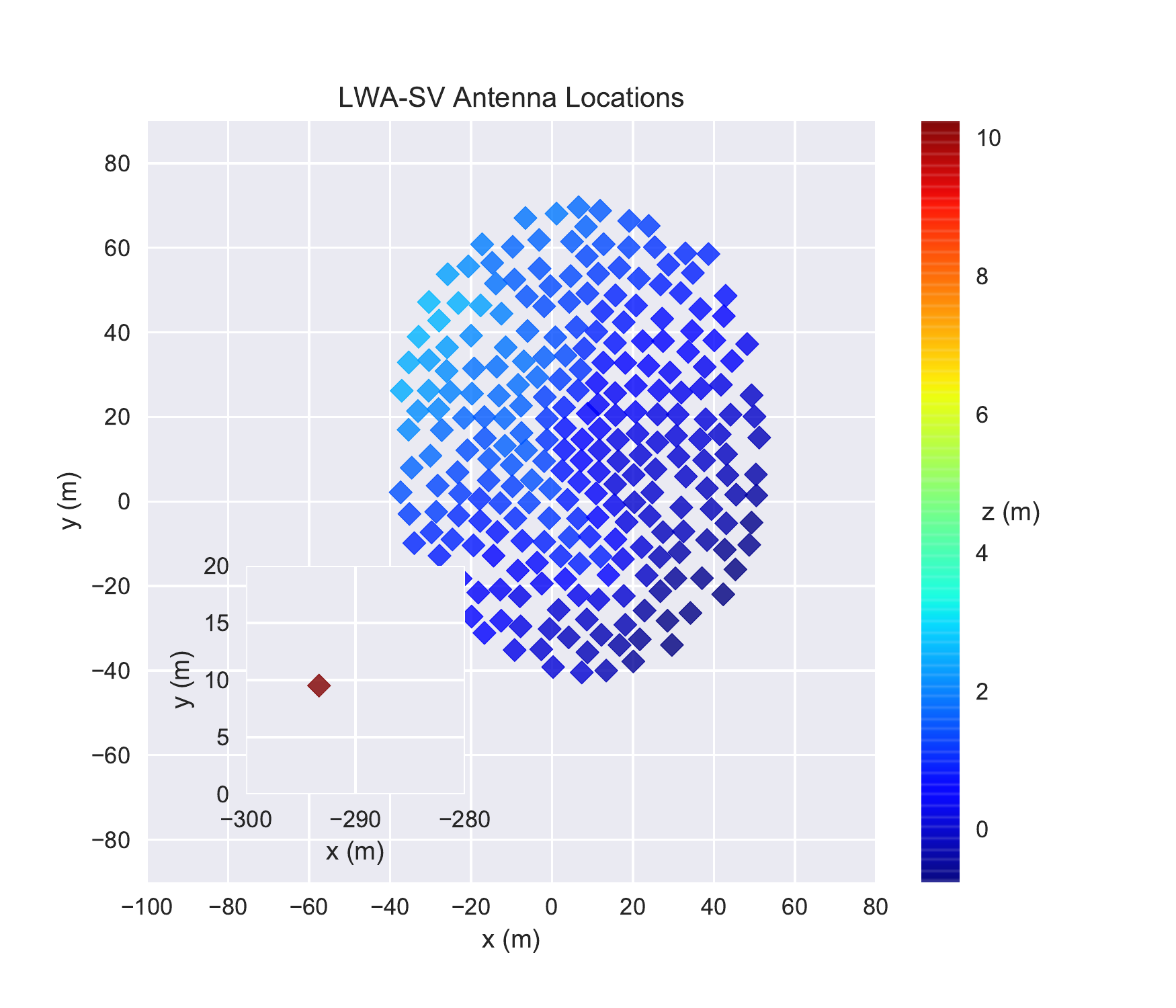}
    \caption{Antenna Locations of the LWA-SV. The w co-ordinate corresponds to the color of each antenna marker. As can be seen LWA-SV has some non-coplanarity within its dense core. The outrigger antenna, inset, is several hundred meters away and adds angular resolution to the interferometer, but is also at a greater $w$ co-ordinate causing non-coplanarity.}
    \label{fig:lwa_locs}
\end{figure}

The LWA-SV antenna locations can be seen in Figure \ref{fig:lwa_locs}, where the color of each antenna 
marker corresponds to its $z$ co-ordinate. The dense 
core is relatively flat with some minor non-coplanarity.
The outrigger antenna, which greatly contributes to the 
overall angular resolution of the array, is several hundred
metres away from the central core, and roughly 10m higher than the rest of the array. 

Originally, the LWA-SV EPIC correlator gridded the electric 
fields directly and then performed an inverse FFT to the 
electric-field sky space, followed by a squaring and 
accumulation operation to form the final image. This 
step was replaced by the DFT
method described in Section \ref{sec:theory}.

\subsection{Validation with test Data}

The DFT matrix was pre-calculated to not have any
time dependence, and imaged sky cosine 
co-ordinates corresponding to a 64x64 grid in $l$,$m$ space. This 
gives values both in and 
out of the celestial sphere, to simplify post-image
rendering but constituting an ``all-sky" image. The antenna 
cable delays and gains were factored into the DFT matrix 
using the LWA Software Library \citep{dowell_long_2012}. 

The data used was captured at 74 MHz as Cygnus A and the Galactic
plane transited overhead. To demonstrate the wide-field
errors that occur without $w$-correction, the DFT was calculated twice: once as 
stated above, and the second time with the $w$-term in 
the antenna locations set to zero. The latter method
simulates imaging with no $w$-correction, similar to the 
original demonstration of the EPIC correlator. In an FFT based
method, correct anti-aliasing of the antenna locations relative
to the Fourier grid points also has to be accounted for, without
which the error will increase further. There is no need to compensate for aliasing with the DFT approach. 

\begin{figure}
    \centering
    \includegraphics[width=\columnwidth]{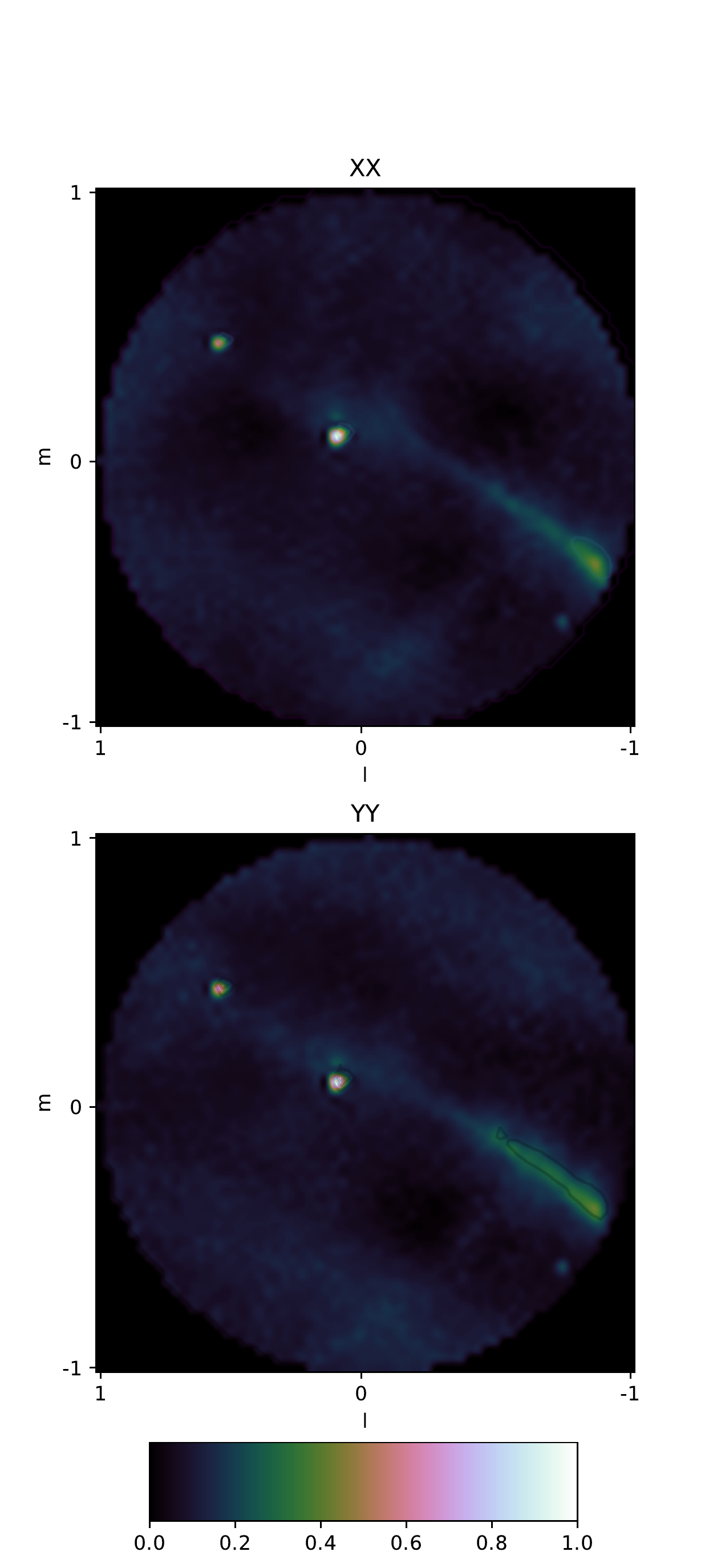}
    \caption{Image, normalised to unity at the peak, formed with the DFT Matrix using the electric
    fields for low frequencies. This offers perfect $w$-correction within numerical precision. Auto-correlations have not been removed in this image, which gives an identical positive offset to all pixels. Cygnus A is the bright pixels in the central region of the image, with Cassiopeia A to towards the bottom left. The Galactic plane is also visible. Units are arbitrary and normalised to a max value of 1.}
    \label{fig:dft_image}
\end{figure}

\begin{figure*}
    \centering
    \includegraphics[width=0.75\textwidth]{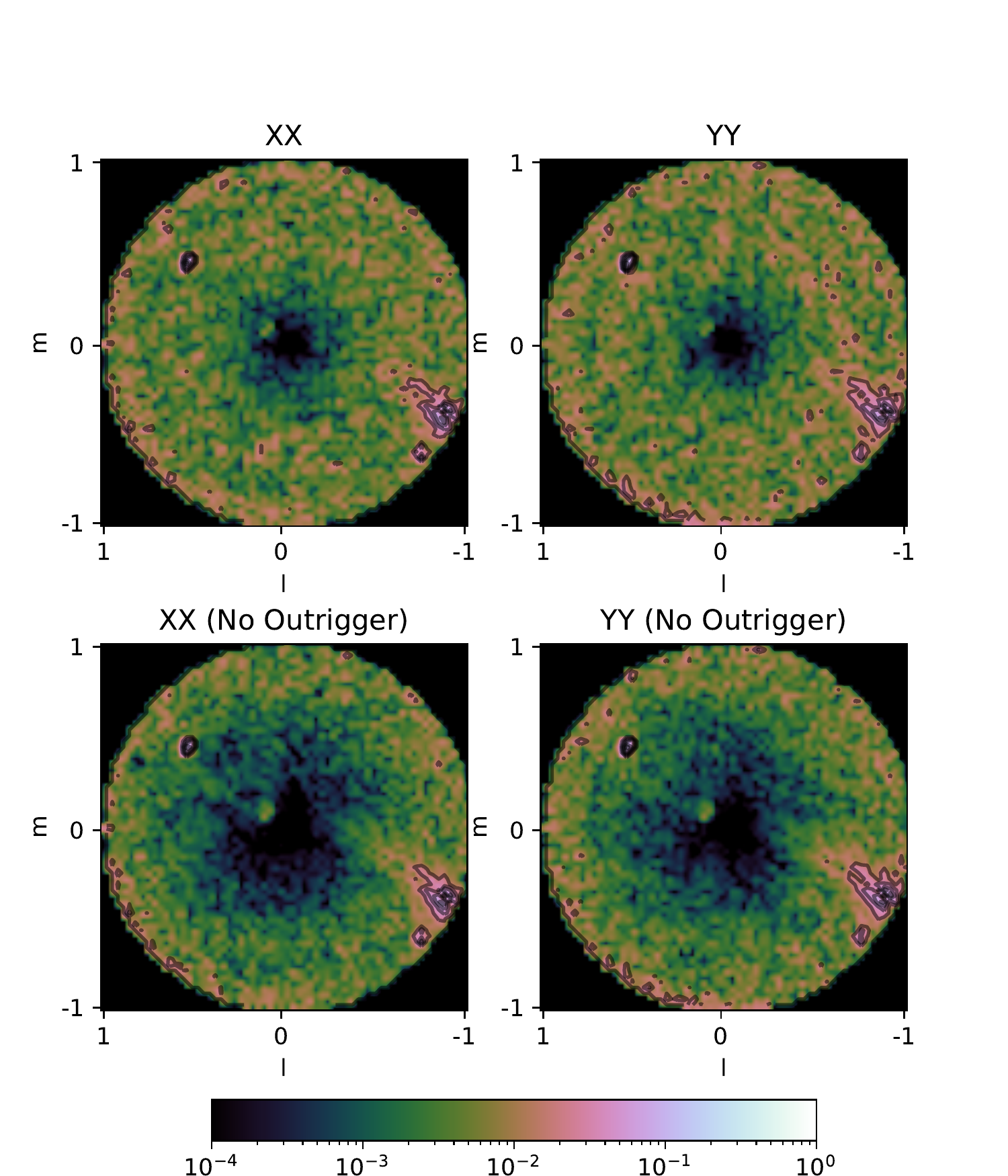}
    \caption{Difference between DFT Image, in Figure \ref{fig:dft_image}, and the DFT Image with no $w$-correction for XX and YY polarisations. This is equivalent to the difference between the DFT image and the image formed with the EPIC architecture where no wide field correction is made. Errors approach 10\% low in the beam, with real astrophysical structure being strongly affected. Contours start at 0 and correspond to levels of 0.2. When the outrigger is removed the wide field error becomes significantly less but still quite substantial, especially around structure such as the Galactic plane.}
    \label{fig:dft_diff}
\end{figure*}

A single channel image of Cygnus A and the Galactic plane is shown in 
Figure \ref{fig:dft_image}. An image difference is shown 
in Figure \ref{fig:dft_diff} where no $w$-correction has 
been performed. The auto-correlations have not been
removed, which add a DC offset to the image. It is important to note that
this imaging is centred on the zenith directly overhead. In arrays with a 
constant slope (such as on a hillside) the effective zenith may
not be directly overhead. The LWA is on such a slope, but still with 
significant non-coplanarity from a calculated best-fit slope. 
A phase correction can be multiplied in corresponding to the geometry of
this slope. In the case of the imaging here with
the LWA this was not applied, thus imaging is centred on the zenith. Regardless,
the majority of the non-coplanarity and wide-field error results from the outrigger.

As can be shown, even for an array such
as LWA-SV, there are significant wide-field errors
on the order of 10\% for the dirty map produced. This is 
surprising, as it was originally thought that for a low frequency
observation with an almost coplanar
telescope such as the LWA site at Sevilletta, wide-field
errors would be insignificant.

\subsection{DFT Performance}

The performance of using the DFT for this low frequency
test case was found to be comparable to the original EPIC
architecture of using a convolutional gridding followed
by an FFT. This is likely due to the matrix multiplication
used for the DFT, with dense matrix multiplications being particularly efficient when implemented
on GPUs. Even with optimised algorithms such as the 
convolution scheme described in \cite{romein_efficient_2012}, and the CUDA FFT library, these
algorithms do not map as well to the GPU model as dense 
matrix multiplications do, and suffer from low performance in comparison as a result. 
The DFT multiplications were batched, with a single multiplication being performed
for each channel and polarisation.

The run-times, averaged over 50 correlation operations, for
each time ``gulp" of data is shown in Table \ref{tab:benchmark_times}. A time gulp in this instance corresponds to the coarse chunks of electric field data which are 
decimated in time that transit through the Bifrost framework. It was observed that a square data matrix, $X$, resulted in the most efficient multiplication times. This is likely due to the optimum benefits this provides in terms of locality, caching and sub-division of the matrix multiplication algorithm using a suitable ``blocking" matrix multiplication algorithm \citep{lam_cache_1991}. The DFT performs comparably to the original
grid and FFT approach with EPIC. For optimum wide-field correction it was possible to 
process 16 Channels of LWA data corresponding to 400 KHz of bandwidth.

\begin{table}
\centering
\begin{tabular}{lll}
$64^2$ Image Size & Processing Time \\ 
8 Channels & DFT & EPIC \\
\hline \hline 
No. Timestamps & & \\
\hline
512 (20ms) & 11.8 ms & 10.7 ms \\
1024 (40ms) & 20 ms & 13.3ms \\
\hline 
16 Channels & DFT & EPIC \\ 
\hline \hline
512 (20ms) & 18.9 ms & 11.621 ms \\
1024 (40ms) & 40 ms & 17.4 ms \\
\end{tabular}
  \caption{DFT Imaging run-times for a $64^2$ image using a Pascal P100 GPU, implemented with the Bifrost framework. Averaged over 50 runs. Bracketed numbers are the amount of time that a ``gulp" of timestamps corresponds to. For a direct imaging pipeline to function normally the processing time must not exceed this figure, or it 
  is incapable of running in real time.}
\label{tab:benchmark_times}
\end{table}

\begin{figure}
    \centering
    \includegraphics[width=1.0\columnwidth]{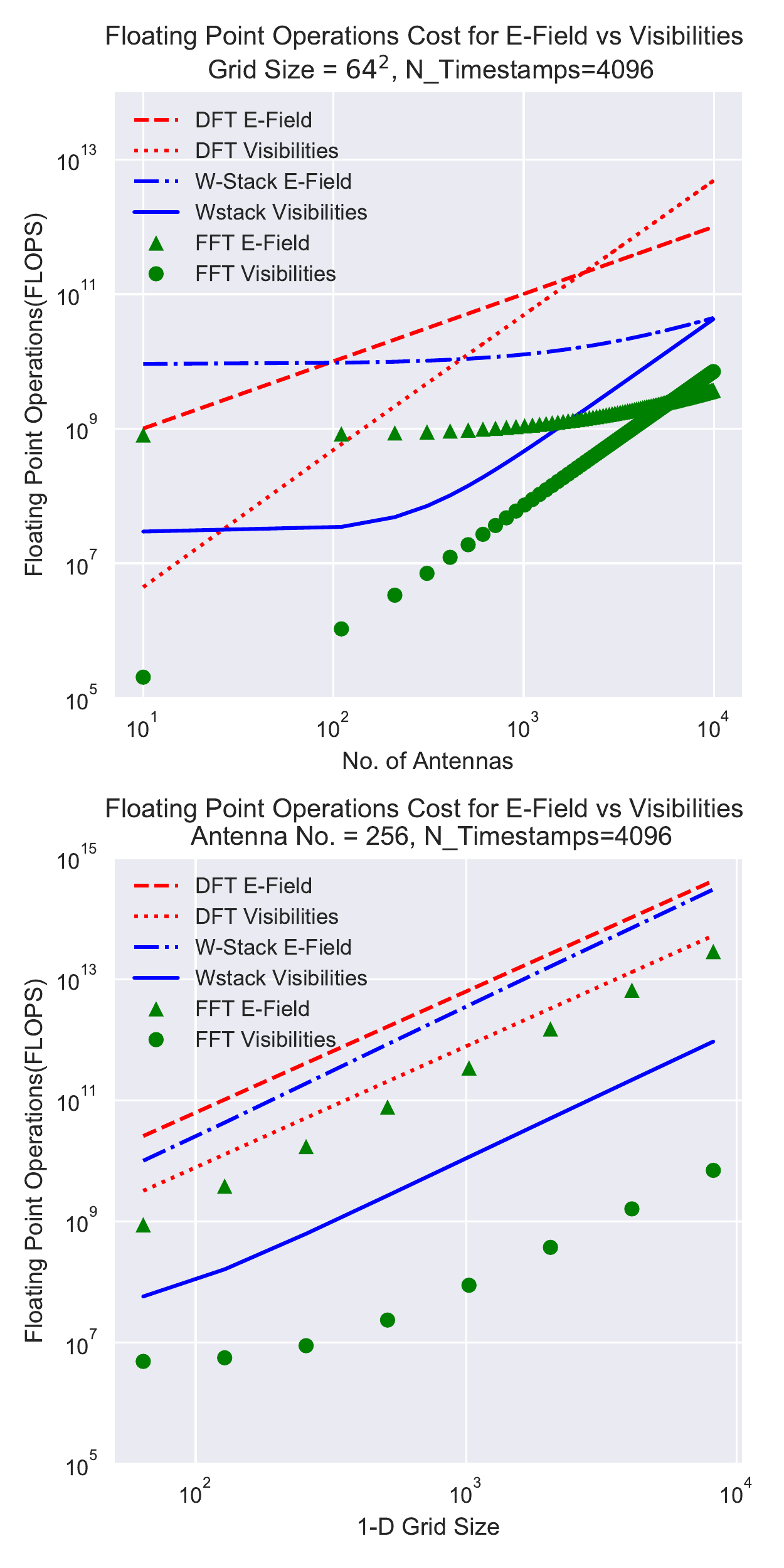}
    %0.5\textwidth]{images/time_gulp_scaling.pdf}
    \caption{Naive floating point costs calculated for different wide field 
    imaging techniques as well as the cost of the ``standard" EPIC 
    architecture using convolutional gridding and an FFT. Whilst the DFT cost is often the highest, algorithmically it lends itself to efficient computational implementation which can negate this in certain parts of the parameter space, and have comparable efficiency to convolutional gridding and FFT based direct imaging.}
    \label{fig:cost_analysis}
\end{figure}

\begin{figure}
    \centering
    \includegraphics[width=\columnwidth]{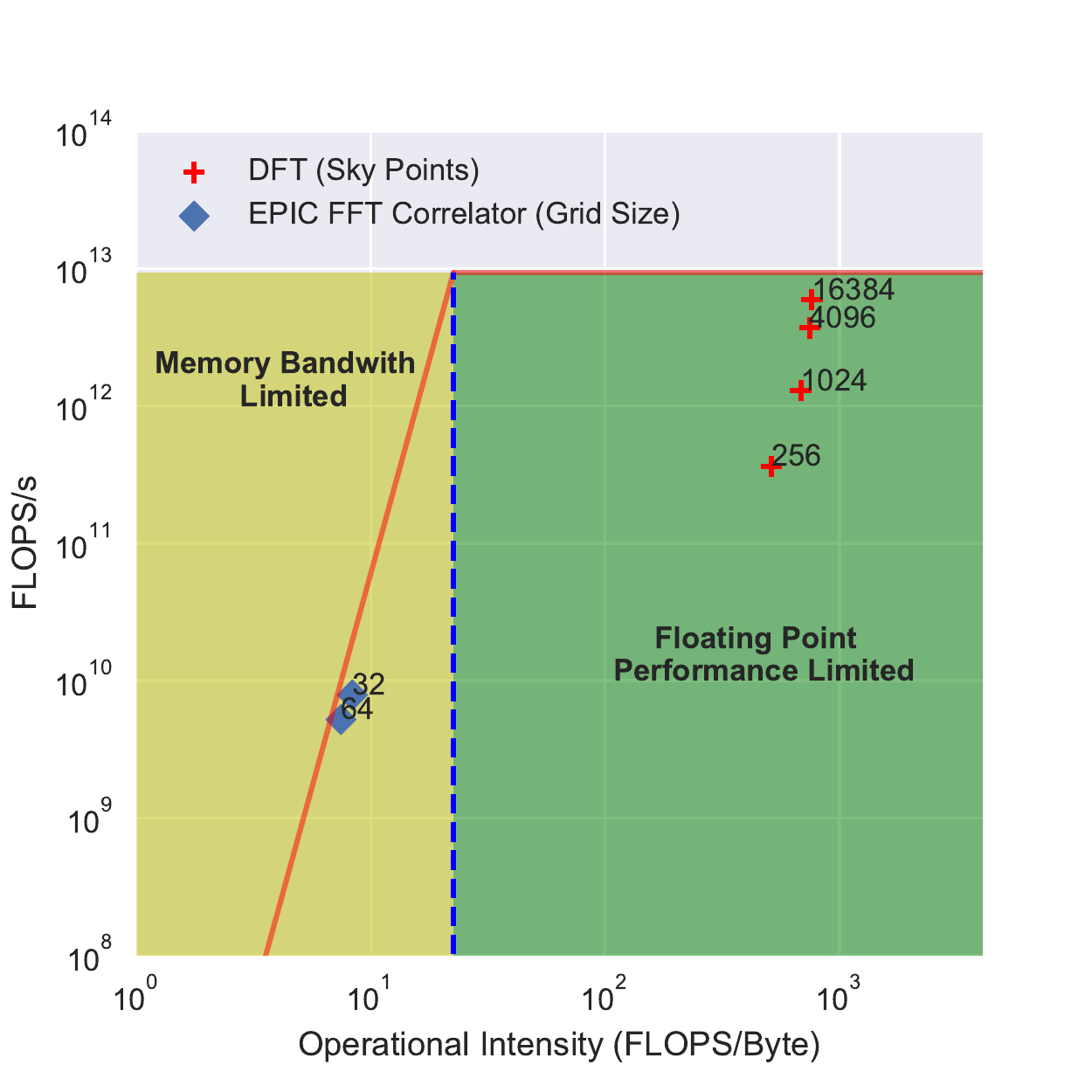}
    \caption{Roofline analysis of the DFT method versus the EPIC FFT pipeline, plotted as Floating Point Operations per Second (FLOPS/s) versus Operational Intensity(FLOPS/Byte). Operational Intensity is a measurement of the number of floating point operations performed per byte loaded from memory. The number next to each point corresponds to the parameter value in brackets in the legend. The roofline plotted is for an NVIDIA Pascal P100 GPU. The sloping red line constitutes the memory bandwidth of the card, and the flat red line the peak floating point performance of the card.  Any algorithm that sits below the sloping roof is fundamentally limited in performance by the available memory bandwidth of the card. Under the flat roof it is only limited theoretically by the peak floating point performance of the card. Its exact placement within the roofline area is dictated by the algorithmic implementation.
    The DFT benefits from much greater operational intensity compared to FFT based methods, which is the reason it runs at comparable speeds despite having inherently more floating point operations.}
    \label{fig:roofline}
\end{figure}

A simulation of the costs associated with Direct Imaging using the 
E-Fields with $w$-stacking versus the DFT or no $w$-correction at all, is shown
in Figure \ref{fig:cost_analysis}. For a small number of antennas, the cost
of $w$-stacking is very inefficient. This is magnified by the low operational 
intensity, which is a measurement of the number of floating point operations per byte of memory loaded, of the gridding and FFT's discussed in Section \ref{sec:wstacking}, with the operational intensity being similar to the EPIC FFT Correlator shown in Figure \ref{fig:roofline}. All methods have roughly squared scaling as a function of the 1-D size of the 2-D grid. 

%and is magnified by low operational intensity, comparable to the EPIC FFT correlator shown in Figure \ref{fig:roofline}, where operational intensity is defined
%as the number of floating point operations performed on each byte loaded from memory, of the algorithms involved. 
%Thus this is likely to be impractical for implementation on a radio interferometer with less than 50 elements. 

The DFT has strong performance in practice, when
Figure \ref{fig:cost_analysis} would lead us to 
believe it slower. To understand this, a roofline
analysis has been performed where the measured computational performance of an algorithm, in floating point operations per second, is plotted versus its operational intensity \citep{williams_roofline:_2009}. In addition the memory bandwidth (the slope) and peak floating point performance (the roof) of the computational architecture on which the algorithm is being run is shown, forming the 'roofline'. This is shown in Figure \ref{fig:roofline}. 

The DFT benefits from significant operational intensity, which in concert with efficient implementations of matrix algorithms such as those noted in \cite{lam_cache_1991} often translates into strong memory locality. This comes with caching benefits, where contiguous elements in use can be stored in memory that is higher speed than main memory RAM, strongly benefiting run-time performance. Thus the algorithmic nature of the DFT is a good fit for modern GPU hardware, leading to higher real world performance than its raw floating point operation cost would lead us to believe. Compared to the original EPIC implementation, the DFT approach replaces both the gridding and FFT. 

As the number of antennas increases however, a convolutional gridding and 
FFT approach may be more efficient provided the layout of the array is 
dense. A denser array allows smaller FFTs using the EPIC formalism  \citep[see][]{morales_enabling_2011,thyagarajan_generic_2017,kent_real-time_2019}. However as shown, it is very challenging to perform wide-field correction in real-time using this approach.

\subsection{DFT Applicability}

The ability to use the DFT matrix as an architecture for accurate direct
imaging is highly array and observation dependent. From the discussed costs and roofline performance figures, the DFT is very
costly from a floating point operations perspective, but lends itself to 
very efficient implementations on current high performance compute hardware. 
Total floating point operations is a poor predictor of performance. 
The parameter that ultimately controls performance and 
applicability is the dimensions of the $D$ matrix. Too many antennas,
sky pixels, or a combination, can make this approach impractical for 
implementation on a real-time direct imaging interferometer. This is due to 
the computation time for a batch of electric field timestamps exceeding the
amount of physical time each batch corresponds to. 

The DFT architecture would be most applicable for a non-redundant and sparse array with many elements, 
where high time resolution imaging is a valuable observation mode. Redundant arrays can use FFT beamforming approaches \citep{masui_algorithms_2019} for reducing data rates, and dense arrays can use the EPIC correlator \cite{thyagarajan_generic_2017,kent_real-time_2019} for direct radio imaging. A combination of these schemes can also be utilised, depending on the interferometer configuration, geometry, and science goals. It is important to remember 
that as  an interferometer becomes more sparse, with implied increase in baseline length,
ionospheric effects become more prominent. Therefore as with any interferometer in this
configuration, correct characterisation of the ionosphere across the array is extremely
important for high fidelity imaging. These can be efficiently incorporated into a time-dependent $D$ matrix. but will increase the complexity of the interferometer.

An overview of current low frequency interferometers which may be applicable for this direct imaging technique are shown in Table \ref{tab:low_frequency_interferometers}. Apart from the LWA, HERA and the MWA are the most suited for the application of a DFT 
direct imaging technique as described here, due to the similar properties in terms of array topology and frequency covered relative to the LWA. They are both at a low frequency with a few hundred antennas, which means that good sky coverage at the angular resolution dictated by the synthesised beam of the array can be achieved. With current consumer
GPU hardware, as used in the LWA-SV correlator, it would likely not be practical to deploy the described technique on CHIME or HIRAX, because of the higher frequencies and greater number of antennas. However with the rapid advances in consumer electronics, in GPUs especially, it may soon be a practical technique. 

In the example shown here, the LWA-SV's resolution makes
it possible to sample the entire sky using the DFT imaging approach
at a resolution representative of the overall angular resolution
of the instrument. If a greater resolution is required, such as 
for higher frequency measurements as with HIRAX or CHIME, 
or greater maximum baseline extent, then the DFT imaging may 
become an impractical choice. In this case,
an EPIC or FX based correlator may be more efficient. Compromises can be 
made by limiting the number of sky pixels sampled, such as to track a particular 
astronomical source. DFT matrix based imaging has the advantage
of being extremely flexible and can fit various interferometers for 
different observation modes.

\begin{table*}
  \centering
  \begin{tabular}{ | P{0.14\linewidth}| P{0.14\linewidth}| P{0.14\linewidth}| P{0.14\linewidth}| P{0.14\linewidth}| P{0.14\linewidth} | }
  \hline
    Telescope & Array Bandwidth & Channel Bandwidth & Antenna Elements & Antenna FoV & Angular Resolution \\
    \hline \hline 
    LWA-SV \citep{taylor_first_2012-1} & 10 MHz - 88 MHz & 25 kHz & 256 Dipole Antennas &  $100^{\circ} $ (74 MHz) & $192^\prime$(74 MHz, Single Station) \\
    \hline
    HERA \citep{deboer_hydrogen_2017} & 50 MHz - 280 MHz & 97.8 kHz & 380 14m Parabolic Dishes & $9^{\circ} $ & $5.14^{\prime}$ - $25.2^{\prime}$ \\
    \hline
    MWA \citep{tingay_murchison_2013,wayth_phase_2018}& 80 - 300 MHz (30.72 MHz processable) & 40 kHz & 256 Tiles of 4x4 Dipole elements & $24.¶^{\circ}$ (150 MHz) - $19.4^{\circ}$ (200 MHz) & $2^\prime$(Precursor) - $3^\prime$(Full Array)  \\
    \hline
    CHIME \citep{stepp_canadian_2014} & 400 MHz - 800 MHz & 390 kHZ & 1280 feeds across 10 cylinders & $100^{\circ} $(EW) $2.5^{\circ} $ - $1.3^{\circ} $(N-S) & $30^\prime$ - $60^\prime$\\ 
    \hline
    HIRAX \citep{newburgh_hirax:_2016} &  400 MHz - 800 MHz & 390 kHz & 1024 6m Parabolic Antennas & $7.48^{\circ} $ - $3.87^{\circ}$ & $5^\prime$ - $10^\prime$ \\
    \hline
    
\end{tabular}
  \caption{An overview of current low frequency radio interfereometers that may be suitable for application of DFT imaging using the electric fields. The LWA-SV (tested), HERA and the MWA are the most applicable targets. HIRAX and CHIME work at higher frequencies necessitating sampling a high number of sky points to form an image representative of the angular resolution of the instrument. They also have more antenna elements, which along with the higher frequency, might make the DFT imaging approach described in this paper practically unfeasible with current compute hardware. In the future, with the continuous advance of consumer electronics, this may change.}
  \label{tab:low_frequency_interferometers}
\end{table*}

With an array of high density, it is more pragmatic to use the
EPIC correlator architecture described in \cite{thyagarajan_generic_2017}, where 
it is more effective to grid the electric fields and then calculate the Fourier
Transform by the FFT algorithm. However the EPIC architecture
is not able to incorporate the outrigger antennas because they lie off
the dense grid. Incorporating them would involve increasing the FFT grid size, 
resulting in a commensurate increase in computation time. The DFT can easily add outrigger antennas to improve instrumental angular resolution through increasing the diameter of the synthesised aperture by adding another column of the $D$ matrix, with linear scaling as a result. 

With this in mind, a direct imaging telescope using the DFT architecture
described here provides accurate wide-field imaging, with
no constraints on the interferometer extent. With the addition of a 
calibration architecture as described in \cite{beardsley_efficient_2017},
it is a highly accurate, fast, and flexible method for direct electric field imaging.

\section{Discussion} 

We have shown that using a Direct Fourier Transform imaging matrix is a tractable solution to the problem of wide-field imaging on direct
electric field imaging interferometers, especially
at low frequencies. This is only 
possible by imaging the electric fields directly from each antenna. A similar approach using visibility data products would be practically unfeasible due to the high computational load.

As shown, standard techniques for wide-field imaging with visibilities are a poor fit to imaging directly with the electric field data from each antenna, due to the requirement for 
them to be done every time step. The compute cost in floating point
operations is made worse by the nature of the algorithms involved. FFTs
perform poorly on GPUs, especially at small sizes, and the difficult to predict memory access pattern of convolutional gridding also decreases efficiency.

The DFT based approach places no restrictions on the placement of antennas or on regularly spaced grid points, as with an FFT. The scaling of this method is $\mathcal{O}(N_K N_T N_A)$, which means adding in additional antennas, sky pixels, or more timestamps causes a linear increase in cost.

Thus the DFT allows real-time high time resolution wide-field imaging, with significantly enhanced flexibility compared to gridding and FFT based methods. The $D$ matrix
described in Section \ref{sec:theory} allows important antenna dependent
terms to be taken into account. Extending the formalism further to include
direction-dependent effects such as ionospheric distortion per antenna is
also mathematically simple, but we do not have access to a suitable instrument to test this with currently.

Using a DFT matrix allows the highly compute bound nature of matrix
multiplication to be taken advantage of, and GPUs are exceptionally 
efficient for compute-bound algorithms of this type. The ability to define
sky points to sample at will opens the possibilities of variable resolutions
across the dirty map. Adding time-dependence to the $D$ operator allows 
calibration terms to be additionally taken into account to create a full featured interferometric imaging framework. 

\section*{Acknowledgements}

Construction of the LWA has been supported by the Office of Naval Research under Contract N00014-07-C-0147 and by the AFOSR. Support for operations and continuing development of the LWA1 is provided by the Air Force Research Laboratory and the National Science Foundation under grants AST-1835400 and AGS-1708855.  A.P.B. is supported by an NSF Astronomy and Astrophysics Postdoctoral Fellowship under award AST-1701440. We gratefully acknowledge the support of NVIDIA Corporation with the donation of a Titan X GPUused for prototyping and testing. GBT and JD acknowledge support from NSF/AST award 1711164. J.K. is funded by Engineering and Physical Sciences Research Council, part of United Kingdom Research and Innovation(UKRI). 

%%%%%%%%%%%%%%%%%%%%%%%%%%%%%%%%%%%%%%%%%%%%%%%%%%


%\FloatBarrier
\bibliographystyle{mnras}
\bibliography{epic_wide}{}
%%%%%%%%%%%%%%%%%%%%%%%%%%%%%%%%%%%%%%%%%%%%%%%%%%

%%%%%%%%%%%%%%%%% APPENDICES %%%%%%%%%%%%%%%%%%%%%

%\appendix

%\section{Noise on an Interferometer at Low Frequency}


%%%%%%%%%%%%%%%%%%%%%%%%%%%%%%%%%%%%%%%%%%%%%%%%%%

% Don't change these lines
\bsp	% typesetting comment
\label{lastpage}
\end{document}